\documentclass[11pt]{amsart}
\usepackage{hyperref}
\theoremstyle{definition}

\theoremstyle{remark}

\numberwithin{equation}{section}

\def \dd{{\rm d}}
\begin{document}

\title[]{Confinement in Einstein's unified field theory}%
\author{Salvatore Antoci}%
\address{Dipartimento di Fisica ``A. Volta'', Pavia, Italy}%
\email{salvatore.antoci@unipv.it}%
\author{Dierck-Ekkehard  Liebscher}%
\address{Astrophysikalisches Institut Potsdam, Potsdam, Germany}%
\email{deliebscher@aip.de}%
\author{Luigi Mihich}%
\address{Dipartimento di Fisica ``A. Volta'', Pavia, Italy}%
\email{Mihich@fisicavolta.unipv.it}


\begin{abstract}
After recalling the mathematical structure of Einstein's Hermitian
extension of the gravitational theory of 1915, the problem,
whe\-ther its field equations should admit of phenomenological
sources at their right-hand sides, and how this addition should be
done, is expounded by relying on a thread of essential insights
and achievements by Schr\"odinger, Kur\c{s}u\-no\u{g}lu,
Lichnerowicz, H\'ely and Borchsenius. When sources are appended to
all the field equations, from the latter and from the contracted
Bianchi identities a sort of gravoelectrodynamics appears, that
totally departs from the so called Einstein-Maxwell theory, since
its constitutive equation, that rules the link between inductions
and fields, is a very complicated differential relation that
allows for a much wider, still practically unexplored range of
possible occurrences.\par In this sort of theory one can allow for
both an electric and a magnetic four-current, which are not a
physically wrong replica of each other, like it would occur if
both these currents were allowed in Maxwell's vacuum. Particular
static exact solutions show that, due to the peculiar constitutive
equation, while electric charges with a pole structure behave
according to Coulomb's law, magnetic charges with a pole structure
interact with forces not depending on their mutual distance. The
latter behaviour was already discovered by Treder in 1957 with an
approximate calculation, while looking for ordinary
electromagnetism in the theory. He also showed that in the
Hermitian theory magnetic charges of unlike sign mutually attract,
hence they are permanently confined entities. The exact solutions
confirm this finding, already interpreted in 1980 by Treder in a
chromodynamic sense.\par The two four-currents considered in this
paper are of pure phenomenological origin. In the last Section a
non phenomenological manner of defining these four-currents is
proposed, that is based on Weyl conformal curvature tensor and on
Petrov eigenvalue equation.

\end{abstract}
\maketitle
\section{Introduction}\label{1}
With their theory of the nonsymmetric field, either in the
metric-affine \cite{Einstein1925, ES1946, Einstein1948, EK1955} or
in the purely affine version \cite{Schroedinger1947a,
Schroedinger1947b, Schroedinger1948, Schroedinger1951}, while
providing a last demonstration of their mathematical insight,
Einstein and Schr\"odinger left as heritage to the future
generations the heavy task of trying to attribute a physical
interpretation to the very similar field equations that, by
proceeding from different startpoints, both of them eventually
arrived at. We shall consider here for definiteness the theory
proposed by Einstein, in its complex, Hermitian version
\cite{Einstein1948}. In this theory, defined on a real,
four-dimensional manifold, one avails, as independent fundamental
quantities, of the Hermitian tensor $g_{ik}=g_{(ik)}+g_{[ik]}$ and
of the Hermitian affine connection
$\Gamma^i_{kl}=\Gamma^i_{(kl)}+\Gamma^i_{[kl]}$. From $g_{ik}$ one
builds the Hermitian contravariant tensor $g^{ik}$ such that
\begin{equation}\label{1.1}
g^{il}g_{kl}=g^{li}g_{lk} =\delta^i_k,
\end{equation}
and, since $g\equiv\det{(g_{ik})}$ is a real quantity, the
Hermitian tensor density
\begin{equation}\label{1.2}
\mathbf{g}^{ik}=\sqrt{-g}g^{ik}.
\end{equation}
In Einstein's Hermitian theory, under quite general conditions
\cite{TH}, the Hermitian affine connection $\Gamma^i_{kl}$ is
uniquely defined by the tensor $g_{ik}$ through the transposition
invariant equation
\begin{equation}\label{1.3}
g_{ik,l}-g_{nk}\Gamma^n_{il}-g_{in}\Gamma^n_{lk}=0.
\end{equation}
Let the further field equation
\begin{equation}\label{1.4}
\mathbf{ g}^{[is]}_{~~,s}=0
\end{equation}
be satisfied. From (\ref{1.3}) one gets \cite{Schroedinger1950}
that (\ref{1.4}) is equivalent to the injunction
\begin{equation}\label{1.5}
\Gamma_i\equiv\Gamma^l_{[il]}=0
\end{equation}
on the skew part of the affine connection. From (\ref{1.3}) alone
it stems further:
\begin{equation}\label{1.6}
\Gamma^a_{(ia),k}=\Gamma^a_{(ka),i}.
\end{equation}
The fulfillment of both (\ref{1.3}) and (\ref{1.4}) is crucial for
the properties of the two generally nonvanishing contractions
$R^i_{~~kli}$ and  $R^i_{~~ilm}$ of the Riemann curvature tensor
\begin{equation}\label{1.7}
R^i_{~~klm}(\Gamma)=\Gamma^i_{kl,m}-\Gamma^i_{km,l}
-\Gamma^i_{al}\Gamma^a_{km}+\Gamma^i_{am}\Gamma^a_{kl}.
\end{equation}
The second contraction reads in general:
\begin{equation}\label{1.8}
R^i_{~~ilm}=\Gamma^i_{il,m}-\Gamma^i_{im,l}.
\end{equation}
When both (\ref{1.3}) and (\ref{1.4}) are satisfied, this second
contraction just vanishes due to (\ref{1.5}) and (\ref{1.6});
hence, like it occurs with the symmetric theory of 1915, also the
problem of choosing which combination of the contractions one
should introduce in the field equations simply disappears. Under
the same circumstances, the first contraction
\begin{equation}\label{1.9}
R_{kl}(\Gamma)=\Gamma^i_{kl,i}-\Gamma^i_{ki,l}
-\Gamma^a_{ki}\Gamma^i_{al}+\Gamma^a_{kl}\Gamma^i_{ai},
\end{equation}
i.e. the Ricci tensor, happens to be Hermitian. Einstein proposed
that its symmetric and skew parts should fulfill the field
equations
\begin{equation}\label{1.10}
R_{(ik)}(\Gamma)=0
\end{equation}
and
\begin{equation}\label{1.11}
R_{[ik],l}(\Gamma)+R_{[kl],i}(\Gamma)+R_{[li],k}(\Gamma)=0
\end{equation}
respectively. The field equations (\ref{1.3}), (\ref{1.4}) and
(\ref{1.10}), (\ref{1.11}) of what Einstein called the Hermitian
generalization of the theory of  gravitation can be deduced from a
variational principle, e.g. in the manner shown by Einstein in
\cite{Einstein1948}, or in the more transparent way, that avails
of the ``starred affinity'', envisaged \cite{Schroedinger1947b} by
Schr\"odinger.\par We have indulged, with these introductory
remarks, in expounding the mathematical structure of Einstein's
Hermitian theory, since the knowledge of the latter is by no means
widespread, while it seems essential for properly understanding
what sort of hopes sustained both Einstein and Schr\"odinger in
their decade-long effort, and what means they believed to be the
most appropriate for trying to fulfill such hopes.\par In the many
technical papers written in the decade 1945-1955 on the subject of
the ``generalized theory of gravitation'', Einstein spent very few
words on the possible physical content of the theory. In his
``Au\-to\-bio\-gra\-phisches'' \cite{Einstein1949} he was very
clear about the reasons for believing that the future progress of
physical theory could not be based on quantum theory, due to the
statistical character of the latter, and to its allowance for the
superposition principle; to him, any real progress could only be
achieved by starting from the general theory of relativity, since
in Einstein's opinion,``its equations are more likely to assert
anything {\it precise} than all the other equations of physics''.
From the discovery of general relativity he had also learned that
no collection of empirical facts, however extensive, could have
been of help in building equations of such intricacy: equations of
such complication can only be retrieved when one has found a
logically simple mathematical condition that determines the
equations in a complete or nearly complete way. Hermitian symmetry
or, more generally, invariance under transposition, that both
represent a natural mathematical extension of the symmetry
properties of the general relativity of 1915, could be
sufficiently strong formal conditions, upon which one might
attempt a generalization of the previous theory, based on real
symmetric quantities.\par At variance with the buoyant optimism
permeating his first attempt on the subject \cite{Einstein1925},
in his later work Einstein, while sometimes asserting that, since
(\ref{1.4}) had to hold everywhere, $g_{[[ik],l]}$ might have to
assume the r\^ole of electric four-current \cite{Einstein1948,
Einstein1955}, became cautious in foretelling what the possible
physical content of his new theory might result to be. In the
autobiographical notes he limited himself to remark that, in his
opinion, equations (\ref{1.3}), (\ref{1.4}), (\ref{1.10}),
(\ref{1.11}) constituted the most natural generalization of the
equations of gravitation, just adding, in a footnote, that in his
opinion the theory had a fair likelihood of proving correct,
provided that the way to a satisfactory representation of the
physical reality on the basis of the continuum will turn out to be
feasible in general. He also believed that, since these equations
constituted the natural completion of the equations of 1915, no
source terms should be appended at their right-hand sides. His
``Au\-to\-bio\-gra\-phisches'' therefore ends with a question
mark, left like a legacy to the posterity: what happens with the
solutions of these equations that are free from singularities in
the whole space?\par On the possible physical content of the
theory, Schr\"odinger was more explicit already in his first paper
\cite{Schroedinger1947a}, where he clearly shows to have perceived
the complete novelty of a fundamental feature of the theory, that
had to become a crucial issue in the years to come, and eventually
led to the abandonment of the efforts aimed at the understanding
of the theory, since it constitutes too large a departure from the
way we are used to think about the electromagnetic interaction.

\section{Interpreting the theory along a path made possible by
Schr\"odinger, Kursunoglu,
Lichnerowicz, H\'ely and Borchsenius}\label{2} In
\cite{Schroedinger1947a} Schr\"odinger deals with his own purely
affine theory, whose field equations, if considered from a
trivially pragmatic standpoint, differ from the ones reported in
Section \ref{1} only due to the presence of the ``cosmological
terms'' $\lambda g_{(ik)}$ and $\lambda g_{[[ik],l]}$ at the
right-hand sides of equations (\ref{1.10}) and (\ref{1.11})
respectively. His remarks about the possible electromagnetic
meaning of his theory can be extended to the case when
$\lambda=0$, and mean that equations (\ref{1.4}) and (\ref{1.11})
should be interpreted like a sort of (modified) Maxwell equations,
with $\mathbf{g}^{[ik]}$ and $R_{[ik]}$ in the r\^oles of
``contravariant density'' and ``covariant field tensor''
respectively. Needless to say, such an interpretation entails a
total departure from the behaviour that one might expect from the
acquaintance with Maxwell's equations in vacuo, where the two
quantities previously mentioned within quotation marks are
mutually related by a simple constitutive equation, that only
entails the metric in the usual tasks of raising indices and
forming densities from tensors. $\mathbf{g}^{[ik]}$ and $R_{[ik]}$
can play in (\ref{1.4}) and (\ref{1.11}) the r\^oles envisaged by
Schr\"odinger only if the constitutive equation of this
``electromagnetism'' is of a kind never heard of before, namely, a
highly involved differential relation, whose content is by no
means surveyable in its explicit form, since its determination
requires first solving (\ref{1.3}) for the affine connection, and
then substituting the resulting expressions
$\Gamma^i_{kl}=\Gamma^i_{kl}(g_{pq}, g_{pq,r})$ in
$R_{[ik]}(\Gamma)$. It is well known \cite{TH} that already the
first step does not yield in general a surveyable outcome, hence
no hint can be drawn a priori about the relation between
inductions and fields dictated by the Hermitian theory.\par
However, despite the total ignorance about its physical meaning,
there is one thing that can be subjected to a close scrutiny in
this sort of electro\-magnetism. In keeping with Schr\"odinger's
and Einstein's conviction that the theory did constitute the
completion of the theory of 1915, no sources are to be allowed at
the right-hand sides of all its field equations. This holds in
particular for (\ref{1.4}) and (\ref{1.11}): as Schr\"odinger
\cite{Schroedinger1947a} notes with some regret, these equations
of unmistakable electromagnetic form are ``used up''; their
left-hand sides cannot be availed of for defining, like it could
have been possible in principle, two conserved four-currents
associated with the skew fields. Therefore, and again at variance
with what occurs in Maxwell's electromagnetism, we have to look
elsewhere for the definition of, say, the electric four-current.
Such a further departure from the known patterns could be welcome
and sought for, because, as complained by Einstein, ``Das Elektron
ist ein Fremder in die Elektrodynamik''. An electric four-current
whose continuous distribution were dictated by the field equations
themselves would represent the solution of many problems that
plague theoretical physics. This is why Einstein suggested that
$g_{[[ik],l]}$ might have to assume the r\^ole of electric
four-current \cite{Einstein1948, Einstein1955}; in
\cite{Schroedinger1947a}
 Schr\"odinger added three more candidates to such a high task.
 But (\ref{1.4}) and (\ref{1.11}) are just the electromagnetic equations that
 one would write {\it in the absence of charges and
 currents} for some continuum endowed with a very strange
 constitutive equation, and the Hermitian theory of relativity is
 a natural generalization of an eminently successful
 predecessor, whose success was however only possible through
 the addition, as source, of the phenomenological energy tensor.
 Therefore the shadow of doubt remained, that the new theory might need phenomenological
 sources too.\par
 Such a doubt was strengthened by the study of the contracted
 Bianchi identities. One may find the derivation of these identities e.g. in
 \cite{Schroedinger1948}, where Schr\"odinger, in keeping with
 his conviction that the theory allowed for a merging of
 gravitational and nongravitational fields in a total entity,
 did not split their expression by separating the terms where
 only symmetric quantities appear
from the terms where only skew quantities occur, like e.g.
Kur\c{s}uno\u{g}lu did a few years later \cite{Kursunoglu1952a,
Kursunoglu1952b}. When the field equations (\ref{1.3}),
(\ref{1.4}) hold, the contracted Bianchi identities found by
Schr\"odinger can be written as
\begin{equation}\label{2.1}
\left[\sqrt{-g}\left(g^{ik}R_{il}+g^{ki}R_{li}\right)\right]_{,k}
=\sqrt{-g}g^{ik}R_{ik,l}.
\end{equation}
 Through the above mentioned splitting, the same identities come to
read
\begin{eqnarray}\label{2.2}
\left(2\sqrt{-g}g^{(ik)}R_{(il)}\right)_{,k}
-\sqrt{-g}g^{(ik)}R_{(ik),l}\\\nonumber
=\sqrt{-g}g^{[ik]}\left(R_{[ik],l}+R_{[kl],i}+R_{[li],k}\right).
\end{eqnarray}
But in \cite{Kursunoglu1952b} Kur\c{s}uno\u{g}lu provided an even
more allusive writing. He noticed that, if one introduces a
symmetric tensor $s^{ik}$ such that
\begin{equation}\label{2.3}
\sqrt{-s}s^{ik}=\sqrt{-g}g^{(ik)},
\end{equation}
where $s$ is the determinant of the tensor $s_{ik}$, and
$s^{ik}s_{il}=\delta^k_l$, the left-hand side of (\ref{2.2}) can
be rewritten as follows:
\begin{eqnarray}\label{2.4}
\left(2\sqrt{-g}g^{(ik)}R_{(il)}\right)_{,k}
-\sqrt{-g}g^{(ik)}R_{(ik),l}\\\nonumber
=\left(2\sqrt{-s}s^{ik}R_{(il)}\right)_{;k}
-\left(\sqrt{-s}s^{ik}R_{(ik)}\right)_{;l}.
\end{eqnarray}
Remarkably enough, the semicolon stands for the covariant
differentiation with respect to the Christoffel symbols built with
$s_{ik}$. Hence the contracted Bianchi identities of Einstein's
nonRiemannian extension of the vacuum general relativity of 1915
admit a sort of Riemannian rewriting that avails of the tensor
$s_{ik}$:
\begin{eqnarray}\label{2.5}
\left(s^{ik}R_{(il)} -\frac12\delta^k_l
s^{pq}R_{(pq)}\right)_{;k}\\\nonumber
=\frac12\sqrt{\frac{g}{s}}g^{[ik]}\left(R_{[ik],l}+R_{[kl],i}+R_{[li],k}\right),
\end{eqnarray}
provided, of course, that equations (\ref{1.3}) and (\ref{1.4})
are satisfied. The same form of the weak identities was arrived at
later \cite{Hely1954a} by H\'ely, who was even more prepared to
appreciate the suggestions coming from  Kur\c{s}uno\u{g}lu's way
of expression, thanks to a precious result \cite{Lichnerowicz1954,
Lichnerowicz1955} found in the meantime: through his study of the
Cauchy problem in Einstein's new theory, Lichnerowicz had
concluded that the metric $l^{ik}$ appearing in the eikonal
equation
\begin{equation}\label{2.6}
l^{ik}\partial_if\partial_kf=0
\end{equation}
for the wave surfaces of the theory had to be
\begin{equation}\label{2.7}
l^{ik}=g^{(ik)},
\end{equation}
or, one must add, any metric conformally related to $g^{(ik)}$.
Since $s_{ik}$, defined by (\ref{2.3}), just belonged to this
class of metrics, H\'ely had one more reason for critically
investigating how the expression (\ref{2.5}) might assume a
physical meaning, like it occurs in the theory of 1915, where the
contracted Bianchi identities just say that the covariant
divergence of the energy tensor is vanishing.\par When confronted
with the weak identity (\ref{2.5}), the sort of regret felt by
Schr\"odinger on noticing that the left-hand sides of (\ref{1.4})
and (\ref{1.11}) were ``used up'' for expressing the vanishing of
two four-currents cannot help becoming a serious concern. One has
to withstand one further disappointment: by adhering to the tenet
endorsed both by Einstein and by Schr\"odinger, according to which
no source terms should be appended at the right-hand sides of
their equations, both sides of (\ref{2.5}) simply vanish. Are we
not missing in this way an occasion offered by the theory? The
very finding of (\ref{2.5}) led Kur\c{s}uno\u{g}lu to modify
\cite{Kursunoglu1952b} Einstein's field equations in order to
provide the weak identities with physical meaning in a field
theoretical way. In a less daring mood, H\'ely appended
\cite{Hely1954b} phenomenological sources at the right-hand sides
of both (\ref{1.10}) and (\ref{1.11}), with the tentative physical
meaning of energy tensor for matter and of electric current
respectively. In such a way, (\ref{2.5}) comes to assert that the
nonvanishing of the covariant divergence of the energy tensor
density of charged matter is due to the Lorentz coupling of its
electric four-current with the electromagnetic field density
$\mathbf{g}^{[ik]}$.\par In the same mood, one may well ask what
hinders appending phenomenological sources to all the field
equations. The question is even more justified, since a class of
exact solutions to the equations of the Hermitian theory has been
found \cite{Antoci1987}, that intrinsically depend on three
coordinates. Solutions belonging to this class appear endowed with
physical meaning when sources are appended at the right-hand sides
of both (\ref{1.11}) and (\ref{1.4}).\par There is indeed one
hindrance, because, as shown in Section \ref{1}, the satisfaction
of (\ref{1.4}) is just  one of the necessary conditions for
getting a Hermitian Ricci tensor. The remedy was found
\cite{Borchsenius1978} by Borchsenius; one needs substituting the
symmetrized Ricci tensor
\begin{equation}\label{2.8}
\bar{R}_{kl}(\Gamma)=\Gamma^i_{kl,i}
-\frac{1}{2}\left(\Gamma^i_{ki,l}+\Gamma^i_{li,k}\right)
-\Gamma^a_{ki}\Gamma^i_{al}+\Gamma^a_{kl}\Gamma^i_{ai},
\end{equation}
for the plain Ricci tensor (\ref{1.9}). The substitution does not
affect the original field equations of Einstein and Schr\"odinger
in vacuo, since there the modified Ricci tensor of Borchsenius is
equal to the true Ricci tensor, but is effective in obtaining a
set of equations with sources that is always Hermitian. When
$s_{ik}$ is adopted as metric, in the footsteps of H\'ely, this
set, whose derivation is reported e.g. in \cite{Antoci1991},
reads:
\begin{equation}\label{2.9}
\mathbf{g}^{qr}_{~,p}+\mathbf{g}^{sr}\Gamma^q_{sp}+\mathbf{g}^{qs}\Gamma^r_{ps}
-\mathbf{g}^{qr}\Gamma^t_{(pt)}
=\frac{4\pi}{3}(\mathbf{j}^q\delta^r_p-\mathbf{j}^r\delta^q_p),
\end{equation}
\begin{equation}\label{2.10}
\mathbf{g}^{[is]}_{~~,s}={4\pi}\mathbf{j}^i,
\end{equation}
\begin{equation}\label{2.11}
\bar{R}_{(ik)}(\Gamma)=8\pi(T_{ik}
-\frac{1}{2}s_{ik}s^{pq}T_{pq}),
\end{equation}
\begin{equation}\label{2.12}
\bar{R}_{[[ik],l]}=8\pi K_{ikl}.
\end{equation}
In this way the two conserved four-currents $\mathbf{j}^i$ and
$K_{ikl}$, and the symmetric energy tensor $T_{ik}$ are appended
to the original equations in a manner that does not spoil their
Hermitian character, and uniquely defines the phenomenological
sources in terms of their geometric counterparts. The relevant
contracted Bianchi identities are \cite{Antoci1991} in this case
\begin{eqnarray}\label{2.13}
-2(\mathbf{g}^{(is)}\bar{R}_{(ik)}(\Gamma))_{,s}
+\mathbf{g}^{(pq)}\bar{R}_{(pq),k}(\Gamma)\\\nonumber
=2\mathbf{g}^{[is]}_{~~,s}\bar{R}_{[ik]}(\Gamma)
+\mathbf{g}^{[is]}\bar{R}_{[[ik],s]}(\Gamma).
\end{eqnarray}
By substituting here the material sources defined above, and by
defining the contravariant energy tensor density as
\begin{equation}\label{2.14}
\mathbf{T}^{ik}=\sqrt{-s}s^{ip}s^{kq}T_{pq},
\end{equation}
one eventually extends H\'ely's result \cite{Hely1954b} to the
form
\begin{equation}\label{2.15}
\mathbf{T}^{ls}_{;s}=\frac{1}{2}s^{lk}
\left(\mathbf{j}^i\bar{R}_{[ki]}(\Gamma)
+K_{iks}\mathbf{g}^{[si]}\right),
\end{equation}
where the semicolon again indicates the covariant derivative done
with respect to the Christoffel connection built with $s_{ik}$. By
completing H\'ely's proposal, this equation asserts that the
covariant divergence of $\mathbf{T}^{ik}$ does not vanish in
general because of the Lorentz coupling of the conserved current
$K_{iks}$ with $\mathbf{g}^{[si]}$, and also because of the
Lorentz coupling of the conserved current density $\mathbf{j}^i$
with the field $\bar{R}_{[ki]}$. But, since the constitutive
equation of this sort of electromagnetism represents a total
departure from the one prevailing in the vacuum of Maxwell's
electromagnetism, we shall not fear that the duality present in
the latter shall lead to a duplicate representation of the same
physical behaviour, with electric and magnetic four-currents both
producing the same phenomena under a duality transformation. In
Maxwell's electromagnetism this occurrence is avoided by imposing,
in keeping with experience, that magnetic four-currents do not
exist. In Einstein's Hermitian theory this injunction is neither
required, nor helpful. The exact solutions show in fact that the
two four-currents give rise to completely different interactions,
both seemingly needed for the description of nature.
\section{The electrostatics of Einstein's Hermitian theory}\label{3}
The simple form of equation (\ref{2.15}) should deceive nobody: it
is evident that the ``particle in field'' imagery, already
misleading in Maxwell's electrodynamics, is totally out of place
both in the general relativity of 1915 and in its Hermitian
extension. From such nonlinear theories, both in exact and in
approximate solutions, as well exhibited \cite{EI1949} in the work
of Einstein and Infeld, one must expect a much subtler link
between structure and motion of the field singularities that one
uses for representing masses and charges. A particular example of
this occurrence is evident \cite{ALM2005} in a solution of the
Hermitian theory, that one cannot help calling electrostatic in
the sense of Coulomb. It can be built by the method reported in
\cite{Antoci1987}; if referred to the coordinates $x^1=x$,
$x^2=y$, $x^3=z$, $x^4=t$, its fundamental tensor $g_{ik}$ reads:

\begin{equation}\label{3.1}
g_{ik}=\left(\begin{array}{rrrr}
 -1 &  0 &  0 & a \\
  0 & -1 &  0 & b \\
  0 &  0 & -1 & c \\
 -a & -b & -c & ~d
\end{array}\right),
\end{equation}
where
\begin{equation}\label{3.2}
d=1+a^2+b^2+c^2,
\end{equation}
and
\begin{equation}\label{3.3}
a=i\chi_{,x}, \ b=i\chi_{,y}, \ c=i\chi_{,z}, \ \ i=\sqrt{-1}, \ \
\chi_{,xx}+\chi_{,yy}+\chi_{,zz}=0.
\end{equation}
The solution is static, and its metric $s_{ik}$ can be written as
\begin{eqnarray}\label{3.4}
s_{ik}=\sqrt{d} \left(\begin{array}{rrrr}
 -1 &  0 &  0 & 0 \\
  0 & -1 &  0 & 0 \\
  0 &  0 & -1 & 0 \\
  0 &  0 &  0 & ~1
\end{array}\right)\\\nonumber
-\frac{1}{\sqrt{d}} \left(\begin{array}{cccc}
 \chi_{,x}\chi_{,x} & \chi_{,x}\chi_{,y} &  \chi_{,x}\chi_{,z} & 0 \\
 \chi_{,x}\chi_{,y} & \chi_{,y}\chi_{,y} &  \chi_{,y}\chi_{,z} & 0 \\
 \chi_{,x}\chi_{,z} & \chi_{,y}\chi_{,z} &  \chi_{,z}\chi_{,z} & 0 \\
 0 & 0 & 0 & 0
\end{array}\right),
\end{eqnarray}
hence the square of the line element, in the adopted coordinates,
reads
\begin{equation}\label{3.5}
\dd s^2=s_{ik}\dd x^{i}\dd x^{k} =-\sqrt{d}\left(\dd x^2+\dd
y^2+\dd z^2-\dd t^2\right) -\frac{1}{\sqrt{d}}(\dd\chi)^2.
\end{equation}
The solution always fulfils the equation $g_{[[ik],l]}=0$, and one
feels entitled to call it electrostatic in the sense of Coulomb.
The reason is simple, and geometric in character. It is discussed
in detail in \cite{ALM2005}, to which the reader is referred. Here
we recall it briefly. If one allows for sources at the right-hand
side of the Laplacian occurring in (\ref{3.3}), one notices that
the admission of such sources in the representative space
corresponds to introducing  a true charge density at the
right-hand side of (\ref{2.10}). Imagine now trying to build
localized true charges by starting from localized, disjoint
sources in the ``Bildraum''. One finds that, when the charges are
very far apart from each other, they will be both pointlike and
spherically symmetric, with all the accuracy needed to account for
the empirical constraints, only provided that the charges occupy,
in the space whose metric is $s_{ik}$, just the positions dictated
by Coulomb's law of electrostatic equilibrium \cite{ALM2005}.\par
One might object that naming ``electrostatic'' the charges
associated with $\mathbf{j}^i$ is wholly premature, since we have
not yet explored what happens when net charges are built from
$K_{ikl}$. But an exact solution allowing for such charges dispels
the objection because, like one might well have expected, the
``magnetostatics'' exhibited by such a solution has nothing to do
with Maxwell's electromagnetism.

\section{In Einstein's Hermitian theory the magnetic
charges are confined entities}\label{4}

One solution of this kind is easily found by the method given in
\cite{Antoci1987}; when referred to polar cylindrical coordinates
$x^1=r$, $x^2=z$, $x^3=\varphi$, $x^4=t$, its fundamental tensor
$g_{ik}$ reads:

\begin{equation}\label{4.1}
g_{ik}=\left(\begin{array}{rrrr}
  -1 & 0 & \delta & 0 \\
  0 & -1 & \varepsilon & 0 \\
  -\delta & -\varepsilon & \zeta & \tau \\
  0 & 0 & -\tau & ~1
\end{array}\right),
\end{equation}
with
\begin{equation}\label{4.2}
\zeta=-r^2+\delta^2+\varepsilon^2-\tau^2,
\end{equation}
and
\begin{equation}\label{4.3}
\delta=ir^2\psi_{,r}, \ \varepsilon=ir^2\psi_{,z}, \
\tau=-ir^2\psi_{,t}, \ \
\psi_{,rr}+\frac{\psi_{,r}}{r}+\psi_{,zz}-\psi_{,tt}=0.
\end{equation}
Its metric $s_{ik}$ can be written as
\begin{eqnarray}\label{4.4}
s_{ik}=\frac{\sqrt{-\zeta}}{r} \left(\begin{array}{rrcr}
 -1 &  0 &   0  & 0 \\
  0 & -1 &   0  & 0 \\
  0 &  0 & -r^2 & 0 \\
  0 &  0 &   0  & 1
\end{array}\right)\\\nonumber
+\frac{r^3}{\sqrt{-\zeta}} \left(\begin{array}{cccc}
 \psi_{,r}\psi_{,r} & \psi_{,r}\psi_{,z} & 0 &  \psi_{,r}\psi_{,t} \\
 \psi_{,r}\psi_{,z} & \psi_{,z}\psi_{,z} & 0 &  \psi_{,z}\psi_{,t} \\
          0         &          0         & 0 &           0         \\
 \psi_{,r}\psi_{,t} & \psi_{,z}\psi_{,t} & 0 &  \psi_{,t}\psi_{,t} \\
 \end{array}\right),
\end{eqnarray}
hence the square of the line element, in the adopted coordinates,
reads
\begin{equation}\label{4.5}
\dd s^2=s_{ik}\dd x^{i}\dd x^{k}
=\frac{\sqrt{-\zeta}}{r}\left(-\dd r^2-\dd z^2-r^2\dd
{\varphi}^2+\dd t^2\right) +\frac{r^3}{\sqrt{-\zeta}}(\dd\psi)^2.
\end{equation}
Let us consider the particular, static solution for which
\begin{equation}\label{4.6}
\psi=-\sum_{q=1}^n K_q\ln\frac{p_q+z-z_q}{r}, \ \
\end{equation}
where
\begin{equation}\label{4.7}
p_q=[r^2+(z-z_q)^2]^{1/2};
\end{equation}
$K_q$ and $z_q$ are constants. One obtains
\begin{eqnarray}\label{4.8}
\delta=i\sum^n_{q=1}\frac{K_q r(z-z_q)}{p_q},\\\nonumber
\varepsilon=-i\sum^n_{q=1}\frac{K_q r^2}{p_q}, \ \ \tau=0,
\end{eqnarray}
and
\begin{equation}\label{4.9}
\zeta=-r^2(1+F),
\end{equation}
with
\begin{eqnarray}\label{4.10}
F=\sum^n_{q=1}K_q^2 +r^2\sum^n_{q=1}~\sum^{n(q'\ne q)}_{q'=1}
\frac{K_qK_{q'}}{p_qp_{q'}}\\\nonumber
+\sum^n_{q=1}\frac{K_q(z-z_q)}{p_q} \ \sum^{n(q'\ne
q)}_{q'=1}\frac{K_{q'}(z-z_{q'})}{p_{q'}}.
\end{eqnarray}
Let $n=1$, $z_1=0$. Then
\begin{equation}\label{4.11}
\delta=i\frac{Krz}{(r^2+z^2)^{1/2}}, \ \
\varepsilon=-i\frac{Kr^2}{(r^2+z^2)^{1/2}}, \ \ \zeta=-r^2(1+K^2),
\end{equation}
and the interval reads
\begin{equation}\label{4.12}
\dd s^2=\sqrt{1+K^2}\left(-\dd r^2-\dd z^2-r^2\dd {\varphi}^2+\dd
t^2\right) +\frac{K^2}{\sqrt{1+K^2}}\frac{(z\dd r-r\dd
z)^2}{r^2+z^2}.
\end{equation}
It is easy to ascertain that this interval displays a constant
deviation from elementary flatness along the $z$-axis. The length
$\dd l$ of an infinitesimal vector $\dd x^i$, lying in a meridian
plane, orthogonal to the $z$-axis, and drawn from a point for
which $r=0$, $z=\text{const.}$, reads
\begin{equation}\label{4.13}
\dd l=\left(-s_{11}+\frac{(s_{12})^2}{s_{22}}\right)^{1/2} \dd
x^1,
\end{equation}
while the length of the infinitesimal circle drawn by the tip of
the vector $\dd x^i$, when it is so rotated around the $z$-axis
that $\varphi$ grows by the amount $2\pi$, is
\begin{equation}\label{4.14}
\Delta l=2\pi\sqrt{-s_{33}}.
\end{equation}
Since for the circle drawn in this way $r=\dd x^1$, the value of
the ratio $\mathcal{R}$ between length and radius of the
elementary circle turns out to be
\begin{equation}\label{4.15}
\mathcal{R}=2\pi\sqrt{1-\frac{{\delta}^2}{r^2}},
\end{equation}
hence, for the present particular case with $n=1$, one obtains
\begin{equation}\label{4.16}
\mathcal{R}=2\pi\sqrt{1+K^2}.
\end{equation}
But, in an axially symmetric solution, a constant deviation from
elementary flatness along the symmetry axis can be removed by
simply modifying the definition of the manifold, since nothing
enforces the original, tentative choice $0<\varphi\leq 2\pi$ for
the coordinate $\varphi$.\par Let us first rewrite the interval
(\ref{4.12}) in spherical polar coordinates $R$, $\vartheta$,
$\varphi$, $t$, obtained by performing, in the meridian planes,
the coordinate transformation
\begin{equation}\label{4.17}
r=R\sin{\vartheta}, \ \ z=R\cos{\vartheta}.
\end{equation}
Then (\ref{4.12}) comes to read
\begin{eqnarray}\label{4.18}
\dd s^2=\sqrt{1+K^2}\left[-\dd R^2-R^2\left(\dd {\vartheta}^2
+{\sin^2{\vartheta}}\dd {\varphi}^2\right)+\dd
t^2\right]\\\nonumber +\frac{K^2}{\sqrt{1+K^2}}R^2{\dd
\vartheta}^2.
\end{eqnarray}
By the coordinate transformation and fixation of the manifold
\begin{equation}\label{4.19}
{\varphi{'}}=\sqrt{1+K^2} {\varphi}, \ \ 0<{\varphi{'}}\leq 2\pi,
\end{equation}
the interval becomes
\begin{eqnarray}\label{4.20}
\dd s^2=\sqrt{1+K^2}\left(-\dd R^2+\dd t^2\right)\\\nonumber
-\frac{R^2}{\sqrt{1+K^2}} \left(\dd {\vartheta}^2
+{\sin^2{\vartheta}}\dd {\varphi{'}}^2\right).
\end{eqnarray}
This manifold, besides displaying elementary flatness everywhere,
with the exception of $R=0$, is spherically symmetric too. One
recognizes, in the $g_{ik}$ associated with it, one particular
case of the spherically symmetric solutions \cite{Papapetrou1948}
found by Papapetrou. For this particular solution $\mathbf{j}^i$
is everywhere vanishing, while this is not the case for $K_{ikl}$.
In fact, let us consider in this manifold a closed spatial
two-surface $\Sigma$, and define the invariant integral
\begin{equation}\label{4.21}
I=-\frac{1}{8\pi i}\int_{\Sigma}\bar{R}_{[ik]}\dd f^{ik},
\end{equation}
where $\dd f^{ik}$ is a surface element of $\Sigma$. The integral
is always vanishing if $\Sigma$ does not surround, say, the origin
$R=0$ of the spatial coordinates $R$, $\vartheta$, $\varphi'$. In
the opposite case one finds
\begin{equation}\label{4.22}
I=\frac{K}{\sqrt{1+K^2}},
\end{equation}
i.e. $K_{ikl}$ exhibits a pole of magnetic charge located at $R=0$
in the representative space, which, according to (\ref{4.20}), is
a point charge in the metric sense too.\par When $n=2$, the
solution defined by (\ref{4.1})-(\ref{4.10}) cannot describe the
field of two isolated poles of magnetic charge, lying on the $z$
axis, whatever the choice of $K_1, K_2$ and of $z_1, z_2$ may be.
This negative outcome happens despite the fact that the integral
(\ref{4.21}) is nonvanishing when it is extended to a closed
spatial two-surface $\Sigma$ surrounding either one or the other
of the above mentioned positions, and otherwise arbitrary, thereby
proving the existence of net charges built with $K_{ikl}$ both at
$r=0$, $z=z_1$ and at $r=0$, $z=z_2$ respectively.\par In fact, at
variance with what happens when $n=1$, the ratio (\ref{4.15})
shows that the deviation from the elementary flatness occurring on
the $z$-axis is only piecewise constant, hence it can not be made
to disappear by an appropriate choice of the manifold. Therefore,
when $n=2$, the solution can not be considered as representing the
field of two isolated bodies, just like it happens, in the general
relativity of 1915, with the Weyl-Levi Civita field for two masses
at rest \cite{Weyl17, Levi-Civita19, BW22}.\par The $n=3$ case is
more fruitful, for, if we choose
\begin{equation}\label{4.23}
K_1=K_3=K, \ K_2=-K, \ \ z_1<z_2<z_3,
\end{equation}
we find that
\begin{equation}\label{4,24}
\lim_{r\rightarrow 0}F=K^2
\end{equation}
along the whole $z$-axis. Therefore the ratio $\mathcal{R}$,
defined by (\ref{4.15}), says that the deviation from elementary
flatness, just like in the case $n=1$, can be eliminated through
the appropriate definition of the manifold, by suitably choosing
the range of $\varphi$.\par Let us remind that, in the
electrostatic case \cite{ALM2005}, we have found that the electric
charges did occupy the positions of equilibrium dictated by
Coulomb's law, provided that the charges built with $\mathbf{j}^i$
were pointlike in the metric sense, and that the metric $s_{ik}$
happened to be spherically symmetric in an infinitesimal
neighbourhood of each charge, with all the accuracy needed to meet
with the empirical facts. Let us study under what conditions the
three aligned magnetic charges happen to enjoy the same geometric
properties.\par An inspection of the metric (\ref{4.4}) for this
solution shows that pointlike charges in the representative space
are always pointlike in the metric sense too. To check for the
spherical symmetry in an infinitesimal neighbourhood of each
charge, we need evaluating the interval $\dd s$, expressed by
(\ref{4.5}), in an infinitesimal neighbourhood of each of the
points located at $r=0$, $z=z_i,$ $i=1,2,3$. One finds that, in
the close proximity to all the points of the $z$-axis, the
interval (\ref{4.5}) can be approximated as
\begin{equation}\label{4.25}
\dd s^2= \sqrt{1+K^2}\left(-\dd r^2-\dd z^2-r^2\dd {\varphi}^2+\dd
t^2\right) +\frac{1}{\sqrt{1+K^2}}(r\dd \psi)^2.
\end{equation}
In the close proximity of the three points mentioned above one can
use the further approximation
\begin{equation}\label{4.26}
\frac{1}{\sqrt{1+K^2}}(r\dd \psi)^2 =\frac{K^2}{\sqrt{1+K^2}}
\frac{\left[(z-z_i)\dd r-r\dd z\right]^2}{r^2+(z-z_i)^2}.
\end{equation}
Therefore, by performing severally, in the meridian planes, the
coordinate transformations
\begin{equation}\label{4.27}
r=R\sin\vartheta, \ z-z_i=R\cos{\vartheta},
\end{equation}
for $i=1, 2$ and $3$, one will find that in each infinitesimal
neighbourhood the interval will always take the same form, given
by (\ref{4.18}), i.e. the very form that holds in the whole space
for the solution with $n=1$. As a consequence, if one performs the
transformation and fixation of the manifold (\ref{4.19}) also in
this case with $n=3$, defined by (\ref{4.23}), one finds that the
interval is spherically symmetric in the infinitesimal
neighbourhood of each of the pointlike magnetic charges.\par The
geometrical conditions on the metric field surrounding the
charges, whose fulfillment\footnote{although with the
approximation expounded in \cite{ALM2005}.}, in the electrostatic
solution of Section \ref{3}, ensures that Coulomb's law is an
outcome of the theory, in the particular solution considered here
are always satisfied exactly, whatever the mutual positions of the
three magnetic charges may be, provided that the order
$z_1<z_2<z_3$ is respected. One therefore draws the physical
conclusion that these aligned magnetic charges by no means behave
like magnetic monopoles would do, if they were allowed for, in
Maxwell's electromagnetism. The indifferent equilibrium of the
three charges exhibited by this magnetostatic solution of the
Hermitian theory is only possible if the interaction of the
charges is independent of their mutual distances.\par One can
object to this conclusion, because the fact that the charges are
both pointlike in the metrical sense, and each endowed with a
spherically symmetric infinitesimal neighbourhood for whatever
choice of $z_1<z_2<z_3$, might well mean that these charges are
not interacting at all. But, as soon as the conditions
(\ref{4.23}) for $K_i$ are not respected, a deviation from
elementary flatness appears on stretches of the $z$-axis, that can
not be made to disappear through the choice of the manifold, just
like it occurs in the solution with $n=2$, and also in the
two-body, static solutions of the general relativity of 1915.
Moreover, approximate calculations done by Treder already
\cite{Treder1957} in 1957 both by the EIH method \cite{EIH1938,
EI1949} and by the test-particle method \cite{Papapetrou1951} of
Papapetrou revealed the existence, in this
gravito-electromagnetism, of a central force between the poles
built with $K_{ikl}$, that does not depend on their mutual
distance, and that, in the Hermitian theory, is attractive when
the poles have charges of opposite sign.\par The same conclusion
can be drawn also with an argument that relies on another exact
solution \cite{Antoci1984} belonging to the class described in
\cite{Antoci1987}. The solution is a Hermitian generalization of
the Curzon metric \cite{Curzon1924}. In the cylindrical
coordinates of its representative space two Curzon masses, located
at $r=0$, $z=z_1$ and $r=0$, $z=z_2$ respectively, are endowed
with point magnetic charges. For fixed $z_1$ and $z_2$, by
choosing appropriately the values of the constants associated with
both the masses and the charges, one succeeds in obtaining that no
deviation from elementary flatness occur along the whole $z$-axis.
One interprets this circumstance as showing that the gravitational
force between the masses is balanced by the force that the
magnetic charges exert on each other. From the weak field limit of
this exact solution, when the gravitational pull reduces to the
Newtonian behaviour, one concludes too that the force between the
magnetic charges is attractive when the charges have opposite
sign, and that it does not depend on their mutual
distance\footnote{In the mentioned paper \cite{Antoci1984}, the
deviation from the elementary flatness was calculated by availing
of $g_{(ik)}$ as metric. The calculation was repeated with the
right metric $s_{ik}$, and has provided just the same result.}. In
1980 Treder interpreted \cite{Treder1980} his findings of 1957 in
a chromodynamic sense.
\section{Conclusion}\label{5}
Talking of conclusions, here and now, sounds ironically premature.
We are still at the very beginnings, since the theory represents
such a total departure from the known paths. Considering
$\mathbf{g}^{[ik]}$ and $R_{[ik]}$ as electromagnetic inductions
and fields, like Schr\"odinger first \cite{Schroedinger1947a}
envisaged sixty years ago, leads to a gravito-electromagnetism
endowed with a range of possibilities so wide and unexplored,
thanks to the intricate differential constitutive relation linking
these quantities, that one might well despair that its content
will ever be unraveled, and proved to be physically meaningful or
not. And yet, thanks to approximate and to exact findings, some
glimpses about the possible content of the theory have appeared
during the lapse of the decades. Besides, of course, Einstein's
gravitation of 1915, the theory appears to contain, according to
particular exact solutions, electric charges that behave as
prescribed by Coulomb's electrostatics \cite{ALM2005}, as well as
magnetic poles that interact with forces not depending on their
mutual distance. When confronted with such outcomes, one can not
help remembering the hopes expressed by Schr\"odinger in the paper
quoted above:
\begin{quotation}
``We may, I think, hold out the prospect, that those skew fields
together, whatever may emerge as the appropriate interpretation,
embrace both the electromagnetic and the nuclear field and their
interplay with each other and with gravitation.''
\end{quotation}
and dare suggesting, on the basis of the admittedly scant, but
unambiguous evidence gathered until now, that the work on this
theory, abandoned so many decades ago, be resumed in the years to
come.\par
\section{Petrov eigenvalue equation allows for two
non phenomenological currents}\label{6} Due to its conceptual
relevance, S. Antoci and D.-E. Liebscher feel urged to add, six
years after the Conclusion was written, the following remark.\par
The previous Sections have given examples of solutions endowed
with physical meaning as soon as the two above mentioned
four-currents $\mathbf{g}^{[is]}_{~~,s}\equiv {4\pi}\mathbf{j}^i$
and $\frac{1}{8\pi}\bar{R}_{[[ik],l]}\equiv{K_{ikl}}$ are allowed
for in a phenomenological way. There is however an opportunity for
defining these four-currents in a non phenomenological manner,
that has been overlooked up to now. Let $C_{pqik}$ be Weyl's
conformal curvature tensor \cite{Goldberg1962} built with H\'ely's
metric $s^{ik}$, and already quoted in equation (34) of
\cite{Antoci1991}. In Einstein's unified field theory Petrov
eigenvalue equation \cite{Petrov1954} is naturally found to read:
\begin{equation}\label{6.1}
\mathbf{g}^{[pq]}C_{pqik}=\lambda\mathbf{g}_{[ik]},
\end{equation}
where $\mathbf{g}_{[ik]}$, in keeping with Schr\"odinger's idea
\cite{Schroedinger1947a}, has the role of generalized
electromagnetic induction, and $\lambda$ is an eigenvalue. If
$\lambda$ is assumed to be a constant, a momentous occurrence
happens. Imagine substituting Petrov eigenvalue equation
(\ref{6.1}) for the sourceless equations (\ref{1.4}) and
(\ref{1.11}) of Einstein's unified field theory. Provided that
$\mathbf{g}_{[ik]}$ is an eigenvector of (\ref{6.1}), and that it
is suitably normalised, ${g}_{ik}$ is defined by a new set of
field equations in which the two conserved four-currents
$\mathbf{g}^{[is]}_{~~,s}\equiv {4\pi}\mathbf{j}^i$ and
$K_{ikl}\equiv\frac{1}{8\pi}\bar{R}_{[[ik],l]}$ are generally
nonvanishing functions of the coordinates. Of course, in the new
set, the affine connection is not given by (\ref{1.3}), but by
(\ref{2.9}) and (\ref{2.10}), while the symmetrized Ricci tensor
$\bar{R}_{ik}$ has the form (\ref{2.8}) envisaged by Borchsenius
\cite{Borchsenius1978}, and its symmetric part fulfills the
equation
\begin{equation}\label{6.2}
\bar{R}_{(ik)}(\Gamma)=0.
\end{equation}
The new set of equations is thereby complete, in the preliminary
sense that the number of equations and the number of unknowns are
equal, but it is defined only when $\mathbf{g}_{[ik]}$ is one of
the eigenvectors of Petrov equation (\ref{6.1}). The worthiness of
the new set is unknown, as well as its link with the original
equations of Einstein's unified field. At present, one does not
know whether the two above defined four-currents tend to vanish at
positions far away  from the sources, and the original equations
are thereby spontaneously recovered. Moreover, one does not know
whether, for the eigenvectors, the two four-currents turn out to
be everywhere regular, thereby lifting an issue that has afflicted
physics since the onset of electrodynamics. We can just invite
theoretical physicists to start working on the new set of
equations and get its solutions. Does the confluence of two lines
of thought, one originated from Weyl, another one from Einstein
and Schr\"odinger, eventually attains a quantum behaviour thanks
to the geniality of the Petrov eigenvalue equation? When comparing
the field equations (\ref{1.4}) and
 (\ref{1.11}) of Einstein's unified field theory with (\ref{6.1})
 one is led to wonder why Einstein and Schr\"odinger relied
 successfully but exclusively on the Ricci part of the Riemann tensor,
 and that maybe the reliance to an equation for the "skew part"
 $\mathbf{g}_{[ik]}$ where the full properties of the Weyl curvature
 tensor are present would have been necessary to cope with the
 several interactions, hierarchies and occurrences that later appeared in
 theoretical physics.
\newpage
\bibliographystyle{amsplain}

\end{document}